\documentstyle[preprint,aps,floats]{revtex}
\begin{document}
\preprint{CERN-TH/96-98}
\title{ Electroweak Baryogenesis and the Expansion Rate of the
Universe}
\author{Michael Joyce}
\address{Theory Division, CERN, 1211 Geneve 23, Switzerland}
\date{4/6/96}
\maketitle

\begin{abstract}
The standard requirement for the production of baryons
at the electroweak phase transition, that the
phase transition be first order and  the
sphaleron bound be satisfied, is predicated on the assumption of
a radiation dominated universe at that epoch.
One simple alternative -
domination by the energy in a kinetic mode of a scalar field
which scales as $1/a^6$ -
gives a significantly weakened sphaleron bound for the preservation
of a baryon asymmetry produced at a first-order phase transition,
and allows the possibility that the observed baryon asymmetry be
produced when the phase transition is second-order or cross-over.
Such a phase of `kination' at the electroweak scale
can occur in various ways as a scalar field
evolves in an exponential potential after inflation .
\\
\end{abstract}

The Hubble expansion rate $H$ of a
homogeneous and isotropic Big Bang universe
is given by the very simple formula
\begin{equation}
H^2= (\frac{\dot{a}}{a})^2= \frac{8 \pi G}{3}\rho - \frac{k}{a^2}
\label{eq: einstein}
\end{equation}
where $a$ is the scale factor, $\rho$ is the energy density
and $k$ is a constant which depends on the spatial
curvature \cite{Peebles}. The main contribution
to $\rho$ today comes from matter which scales as $1/a^3$,
with perhaps also a curvature term and even
a small cosmological ($\rho=$constant) term.
Going back in time the scale factor decreases and
the energy density in the microwave background radiation blueshifts,
scaling
as $1/a^4$, until it comes to dominate
the right handside of (\ref{eq: einstein}).
The most impressive evidence for this extrapolation
comes from nucleosynthesis.
The precise abundances
of the various nuclei synthesised from the
nucleons as the universe cools below $\sim 1 MeV$
depends sensitively on the relation between the temperature of the
radiation (which goes as $1/a$) and the expansion rate,
and the radiation dominated
picture does remarkably well.

Going back further in
time we reach the electroweak epoch at $T\sim 100 GeV$.
The expansion rate
again enters in determining the details of the relics
left behind, most notably the baryon asymmetry \cite{Mishareview}.
In this Letter it is pointed out that relaxing the standard
assumption of radiation domination at the electroweak
scale has important consequences for electroweak baryogenesis.
The fact that the sphaleron bound and
the usually assumed impossibility of baryogenesis at
a second order or cross-over phase transition
are highly dependent on this assumption
is illustrated with the example of a universe dominated
by the energy in a kinetic mode of a scalar field.
Other examples of alternatives to radiation domination before
nucleosynthesis have been discussed in works
of Barrow \cite{barrow} and Kamionkowski and Turner
\cite{kamturner}, which consider how
the relic abundances of dark-matter particles
are changed in such scenarios.

Consider first the dynamics of a real scalar field $\phi$
with potential $V(\phi)$. Variation of the action
\begin{equation}
{\cal{S}}= \int d^4 x \sqrt{-g} \left( \frac{1}{2} g^{\mu
\nu}(\partial_{\mu} \phi)^{\dagger} (\partial_{\nu} \phi)
- V(\phi)\right).
\end{equation}
taking the FRW metric with scale factor $a(t)$, gives the
equation of motion for the homogeneous modes, which
can be written
\begin{eqnarray}
\frac{d}{dt}(\frac{1}{2} \dot{\phi}^2 + V(\phi)) + 3 H \dot{\phi}^2=
0
\label{eq: eompotl}
\end{eqnarray}
after multiplication by $\dot{\phi}$.
Defining $\eta(t) = \frac{ V(\phi)}{\frac{1}{2} \dot{\phi}^2}$ and
writing the energy density $\rho(t)=\frac{1}{2} \dot{\phi}^2 +
V(\phi)$, we find
\begin{eqnarray}
\rho(t) = \rho(t_o) e^{-\int_{t_o}^{t} \frac{6}{1 + \eta(t)} H(t) dt}
= \rho(t_o) e^{-\int_{a_o}^{a} \frac{6}{1 + \eta(a)} \frac {da}{a}}
\label{eq: scaling}
\end{eqnarray}
When the kinetic energy dominates
$\eta \rightarrow 0$
and
\begin{equation}
 \rho  \propto \frac{1}{a^6}
\label{eq: sixscaling}
\end{equation}
This represents
the opposite limit to inflation 
driven by the potential
energy with $\eta \rightarrow \infty$ and $\rho(t) \approx
\rho(t_o)$.
Indeed for any homogeneous mode (assuming only that $V(\phi)$
is positive) we have that
\begin{equation}
\rho(t_o)(\frac{a_o}{a})^6 \leq \rho(t) \leq  \rho(t_o) \qquad t \geq
t_o
\label{eq: scalinglimits}
\end{equation}
Putting these limiting behaviours of the
energy density into (\ref{eq: einstein})
one finds $a \propto t^{\frac{1}{3}}$ (with $k=0$)
for the $1/a^6$ scaling, in contrast to
$a \propto e^{Ht}$ for inflation(H=const).
Instead of superluminal expansion in
inflation a  kinetic energy dominated mode of a
scalar potential drives a subluminal expansion
very similar to that of radiation ($a \propto t^{\frac{1}{2}}$)
or matter ($a \propto t^{\frac{2}{3}}$).
Writing the stress energy tensor in terms
of a pressure $p$ and the energy density in
the standard way, the equation of state is
$p=\rho$ for the kinetic mode
in contrast to $p=\frac{1}{3}\rho$(radiation),
$p=0$(matter) and $p=-\rho$(inflation).
I will use the
term {\it kination} to refer to a phase of the universe
dominated by the kinetic energy of a scalar field.
The  `deflationary' universe of \cite{spokoiny}
which will be discussed below  is a
particular example of this, in which the inflaton
evolves into such a kinetic mode \cite{footnoteterms}.

Now let us suppose that an unknown amount of energy is
stored in such a mode at the electroweak
epoch. The expansion rate in (\ref{eq: einstein}) becomes
\begin{equation}
H^2= (\frac{\dot{a}}{a})^2= \frac{8 \pi
G}{3}\frac{\rho_e}{2}((\frac{a_e}{a})^6 + f(a)(\frac{a_e}{a})^4)
\label{eq: einsteinb}
\end{equation}
where  $a_{e}$ is the scale factor when the density
in the mode becomes equal to that in radiation and
$\rho_e$ is the energy density at that time. The factor
$f(a)$ accounts for the effect of decouplings,
and in the approximation that they are instantaneous
is $f(a)=(g(a_e)/g(a))^{\frac{1}{3}}$ where $g(a)$
is the number of relativistic degrees of freedom.
The sphaleron bound \cite{sphalbound}
results from the requirement that
the rate of baryon number violating (sphaleron) processes after
the electroweak phase transition be less than the
expansion rate of the universe so that
the baryon asymmetry (putatively) created at the
electroweak phase transition be ``frozen in''. Thus
\begin{equation}
\Gamma_{sph}\sim T_we^{-E_{sph}/T_w} < H_{6}=  (\frac{H_6}{H_4})
H_{4}
\end{equation}
where $H_{6}$ is the Hubble expansion rate and $H_4=1.66 \sqrt{g_w}
\frac{T^2}{M_pl}$ is the expansion rate
we get if we assume radiation domination in the usual way,
with $g_w$=$g(a_w) \sim 100$ and $T_w\sim100 GeV$.
The bound
on  $E_{sph}$, the sphaleron energy, can thus be written
in terms of the usual bound on the same quantity
$E_{sph}^o$ as
\begin{equation}
E_{sph}=E_{sph}^o - T_w \ln [(\frac{g_w}{g_e})^{\frac{1}{2}}
\frac{T_w}{T_{e}}].
\label{newbound}
\end{equation}
This follows since
$\frac{H_6}{H_4} \approx \frac{1}{\sqrt{f(a_w)}}\frac{a_e}{a_w}$
and $T a$=$f(a) T_e a_{e}$,
where $T_{e}$ is the temperature at
radiation-kinetic energy equality
(at $a=a_e$).

Let us take the following approximate bound from
nucleosynthesis: We allow $10 \%$ of the energy to come from
the coherent mode at
$\sim 1 MeV$, just before the first stage of $n-p$
freezout begins \cite{fkot}. Then $T_{e} \sim 3 MeV$, so taking
$T_w\sim100 GeV$, the bound on the sphaleron energy
is reduced by approximately one quarter from its
usual value of $\sim 45T$ \cite{Mishareview}.
The lower bound on $E_{sph}$
can be translated into constraints
on the parameters in the zero temperature theory, most
notably an upper bound on the lightest Higgs particle.
Constraints are usually derived using the bound
expressed as the ratio of the VEV $v$ in the nucleated bubbles
to the nucleation temperature $T_b$, to which the
sphaleron energy is linearly proportional. Typically
therefore the sphaleron bound will be weakened as
\begin{equation}
\frac{v}{T_b} > 1 \qquad \rightarrow \qquad \frac{v}{T_b} > 0.75
\label{eq: newboundvoverT}
\end{equation}
How significant a difference is this?
According to recent lattice studies of
the electroweak phase transition in
the minimal standard model \cite{kajetal}, \cite{Mishareview},
the `usual' sphaleron bound cannot be satisfied for
{\it any} physical Higgs mass, for a top quark
mass of $m_t=175 GeV$. The `new' bound
in (\ref{eq: newboundvoverT})
is satisfied for Higgs masses up to about
$35GeV$. For $m_t=155 GeV$
the bound changes from about $35 GeV$ (for
the `usual' case) to $50 GeV$. The `new'
bounds are still however too low
to be consistent with the LEP bounds
on the standard model Higgs mass $m_H >65 GeV$.

In extensions of the standard model, such as the
minimal supersymmetric model (MSSM), recent
perturbative \cite{cqw} and non-perturbative
\cite{jckk} analyses indicate that the usual
sphaleron bound can be satisfied in various parts
of experimentally allowed parameter space.
The new bound simply widens this allowed parameter
space. In what sense can this widening be said
to be significant or not? For baryogenesis what
one must calculate given any
set of physical parameters (ultimately to be
fixed by particle physical experiments, we hope)
is a depletion
factor $X$, where $B_f= e^{-X} B(T_o)$
is the baryon number at nucleosynthesis and
$B(T_o)$ is the
baryon number created during the departure from
equilibrium at some temperature $T_o$ (usually
very close to the critical temperature for the
phase transition). It is simple to show that
\begin{equation}
X = \int_{t_o}^{\infty} dt \Gamma_{sph}(t)= H_o^{-1} \int_{0}^{T_o}dT
\frac{\Gamma_{sph}}{T}(\frac{T_o}{T})^p
\end{equation}
where $p=2$ in the case of radiation domination, and
$p=3$ for kination. The extra power in the integral
is negligible because the integral is cut-off
very rapidly due to the exponential dependence
in the sphaleron rate, so that the depletion
factor is simply changed in inverse proportion to the
expansion rate at the phase transition $H_o$.
 The estimate given above allowing for the
potential contribution of the kinetic mode
corresponds to a change in the expansion rate
by up to a factor of $10^5$ (the factor inside the
logarithm in (\ref{newbound})),  so that it could
make  the difference in a given model between
an asymmetry consistent with observation, and
one $e^{-10^5}$ times smaller. This is certainly
in an absolute sense a significant difference!

Has such a change to the expansion rate other
consequences? An expansion rate at the electroweak
scale of $\sim 10^{-11}T$, instead of
$\sim 10^{-16}T$ in the radiation dominated case,
leaves the usual treatment
of the phase transition intact, because the timescale
for the expansion is still very
long compared to thermalization time scales.
Details will change. The phase transition
will proceed slightly differently
e.g. with more supercooling before the nucleation of bubbles
\cite{footnotenucl}.
The slowest perturbative processes, those flipping the
chirality of electrons which have a rate
$\sim 10^{-12}T$, will remain out of equilibrium
leading to minor alterations to various calculations
of baryon number.

With increasing  Higgs masses the phase
transition becomes more weakly first order,
and, according to recent non-perturbative
lattice results \cite{kajetalcrossover}
eventually (at $m_H \approx 80 GeV$ in the standard
model) the line of first order transitions
ends in a second order transition and becomes cross-over.
This means that there is actually no phase transition,
all gauge-invariant observables evolving
continuously as a function of temperature.
In this  case
it has been assumed that a baryon asymmetry
of the observed magnitude cannot be
created,  because the departure from equilibrium
required by the Sakharov conditions is too small,
being controlled  by the expansion rate of the universe
rather than by the much shorter time-scales characterizing the
propagation of bubbles at a first order phase transition
\cite{krs},\cite{imposs2nd}, \cite{brand2nd}.
At a first order phase transiton  too weak to
satisfy the sphaleron bound the same will be true
as, after the completion of the phase transition, the
expansion rate again becomes the relevant timescale.
A very simple calculation
of the baryon asymmetry is possible in these cases
with the assumption of homogeneity in the
evolution of the fields. In various extensions
of the standard model with extra CP violation
there are terms in the
effective action which act like chemical potentials
either for baryon number \cite{TZandMSTV}, \cite{DHSS}
or hypercharge \cite{CKNspon}. In the presence of these
source terms one finds (calculating the equilibrium
with the appropriate constraints) the baryon to entropy ratio
\cite{kinationmodels}
\begin{equation}
\frac{n_B}{s} \sim \frac{H_f}{T_f} \frac{1}{g_w} T_f
\frac{d\theta_{CP}}{dT}|_f
\label{bau2nd}
\end{equation}
where $\theta_{CP}$ is the (dimensionless)
CP violating field during its evolution (times
some  model-dependent suppression) and
the derivative its rate of change when
the asymmetry freezes out at temperature
$T_f$, when the expansion rate is
$H_f$.
When the universe is in a phase of
kination, $H \propto \frac{1}{a^3} \propto T^3$,
so that, taking the estimate above,
we can have $\frac{H_f}{T_f} \sim 10^{-11} (T_f/100 GeV)^2$.
To evaluate the remaining factor exactly
would require a full study of the
detailed dynamics of the phase transition,
which in this case is still well beyond
current capabilities. An examination
of the data available on the
models studied in this regime \cite{kajetalcrossover}
indicates that this factor could be as large as order
one since $\Delta T$, the temperature range
which characterizes the change in the
quantity $\theta_{CP}$ by order one
could potentially be smaller
than $T_f$ by enough to cancel $g_w \sim 100$
- the transition is continuous but `sharp'
(it is only because it is that it makes
sense to talk of a `transition' at all).
It also takes place at
higher temperatures ($200-300 GeV$ in the standard model)
than when the transition is first order ($T_f \sim 100 GeV$).
It is thus possible that
an asymmetry compatible
with the observed $\frac{n_B}{s} \sim 10^{-11}$
could result when the electroweak phase transition
occurs during a phase of kination
which ends just before nucleosynthesis.

The simple but important point is that
the standard arguments which are used to rule  out
the possibility of baryogenesis at the electroweak scale
in many models are predicated on the assumption of
knowledge of the expansion rate.
In fact the one variable in an {\it ab initio} calculation
of electroweak baryogenesis which we cannot access (at least
in principle) through direct measurement is the expansion rate
at the electroweak epoch.
Methodologically it is thus more sensible to ask what expansion rate
would be required to generate the observed asymmetry in any
particular model. That there is  any such expansion rate is
itself a very non-trivial requirement of a theory.
We have just seen that allowing for
the contribution of a kinetic energy dominated scalar
mode opens up the possibility of the creation of
the observed baryon asymmetry at a second-order or
cross-over phase transition. Several other possibilities
have been discussed by the authors of
\cite{barrow} and \cite{kamturner}, in the context of
their consideration of the dependence of the relic
abundances of dark matter particles on the expansion rate.
Barrow considers the case of an anisotropic universe
and Kamionkowski and Turner this and various others
including a Brans-Dicke theory of gravity with
the scalar dominated by its kinetic energy.
In these cases  the net effect
is essentially described by an additional
contribution to the energy density scaling as
$1/a^6$ just like that we have considered.
Beyond these there is the possibility of other
non-standard theories of gravity such as
scalar-tensor theories in which the gravitational
constant varies. The rest of this Letter will
concentrate on the specific model
of domination by the kinetic mode of a scalar
field. It is minimal in the sense that it
sticks to standard Einstein gravity, and is
compatible with the inflationary explanation
of the homogeneity and isotropy of the observed
universe.

What one requires in this case  is
that the energy in the kinetic mode be much
greater than the energy in radiation at the
electroweak scale. An explanation of the
`usual' scenario in which the universe is dominated
by uniform radiation at the electroweak epoch is provided
by inflation: A scalar field $\phi$ displaced from its
minimum rolls in its potential $V(\phi)$, sufficiently slowly
that it satisfies the condition $V(\phi) >> \dot{\phi}^2$
for long enough to inflate a small uniform region
outside our present horizon; the field eventually
reaches its minimum and oscillates about it,
until it decays to produce radiation at the
`re-heat' temperature  $T_{RH}$.
An alternative mechanism for reheating was given
by Spokoiny in \cite{spokoiny}. Instead
of rolling into a minimum and oscillating, the
inflaton rolls in a potential (described below)
so that a period of domination by
its kinetic energy follows inflation, with the resultant
$1/a^6$ scaling discussed above. The universe is
reheated simply by particle production in the
expanding universe, which is proportional to
$H^4$ (for scalar particles nonconformally coupled to
gravity). The requirement that this radiation come
to dominate before nucleosynthesis requires that
the transition from inflation to kination occur
at at a sufficiently large expansion rate,
$H > 10^9 GeV$.  Taking the created particles
to be Higgs bosons, the temperature at which thermalization
occurs is estimated in \cite{spokoiny} to be
$\sim 10^6 GeV$ for the case that the transition
radiation domination occurs just before nucleosynthesis.
This `deflationary universe' therefore corresponds exactly
to what was required in the analysis above: a universe in which there
is thermalized radiation by the electroweak scale but
which is dominated by a coherent kinetic mode
potentially
until just before nucleosynthesis.

To see that this domination by a kinetic mode
over radiation can come about also in
conjunction with the standard reheating scenario,
we consider more carefully the sorts of potential
which are required.
The equations governing the dynamics
of the scalar field  are
\begin{eqnarray}
\ddot{\phi} + 3 H \dot{\phi} +
V'(\phi)=\frac{1}{a^3}\frac{d}{dt}(a^3\dot{\phi}) + V'(\phi)=0
\label{eq: potleoma}\\
H^2 = \frac{1}{3M_{p}^2}(\frac{1}{2}\dot{\phi}^2 + V(\phi))
\label{eq: potleomb}
\end{eqnarray}
where $M_p=1/\sqrt{8 \pi G}$ is the reduced Planck mass,
and we neglect the radiation density assuming the scalar
field energy to dominate.
It is shown in \cite{jhalliwell} that there are
particular attractor solutions to  (\ref{eq: potleoma}) and
(\ref{eq: potleomb}) for the
potential $V(\phi) = V_o e^{- \lambda{\phi/M_p}}$:
\begin{equation}
\phi(t)= M_p \sqrt{2A} \ln (M_pt) \quad a \propto t^A
\quad \frac{V(\phi)}{ \frac{1}{2} \dot{\phi}^2 }=3A-1
\label{eq: jhsoln}
\end{equation}
where $\lambda= \sqrt{{2}/{A}}$
and the origin of $\phi$ is redefined so that
$V_o= M_{p}^4 A(3A-1)$.
{}From (\ref{eq: scaling})
it follows that $\rho \propto 1/a^{\frac{2}{A}}$
($\eta=3A-1$). Values of $A>1$
give power-law inflationary solutions and in the
limit $A \rightarrow \frac{1}{3}$, in which the kinetic energy
dominates, we get the scaling associated with kination.
It is easy to see that potentials steeper
than this will generically have kinetic energy dominated modes
by examining the
solution to (\ref{eq: potleoma}) and (\ref{eq: potleomb})
with $V=0$:
\begin{equation}
\dot\phi(t)= \dot\phi_o (\frac{a_o}{a})^3=\dot\phi_o (\frac{t_o}{t})
\qquad \phi(t)= \phi_o +\dot\phi_o t_o \ln \frac {t}{t_o}
\label{eq: phikin}
\end{equation}
In any potential decreasing faster than the exponential
with $A=\frac{1}{3}$ the potential terms in (\ref{eq: potleoma})
and (\ref{eq: potleomb}) once smaller will decrease faster than
the other terms, and the field will approach a solution
of the form (\ref{eq: phikin}).
Exponential potentials are particularly interesting
because they occur generically in
theories which are compactified,
such as supergravity theories or string theories.

In his analysis Spokoiny realizes the transition from
inflation to kination referred to by taking an exponential
potential where $\lambda$ varies in the appropriate
way. If we suppose instead
that the universe goes through a period of inflation driven
by some other field and reheats in the `usual' way
(by oscillation and decay) leaving the radiation
dominant over whatever energy density
is in the exponential potential, it is simple
to see (adding the contribution of the radiation
to (\ref{eq: potleomb})) that the field begins to roll
when the energies become comparable. If the
exponential again has a $\lambda$ varying in the
appropriate way a period of inflation which cools
the radiation can occur followed by a roll of
the field into a deflationary mode as the
exponential becomes steeper.
Alternatively,
one can consider a potential like $V_oe^{-\phi^2/M_p^2}$
with $\phi_{RH} \approx 0$ (the value of the field
at the end of reheating). A period of
inflation (number of e-foldings $\sim \ln M_p/\phi_{RH}$)
can occur when the potential
energy in the field comes to dominate.
These and other models will be discussed in
more detail in a forthcoming paper \cite{kinationmodels}.

{\it Acknowledgements.\/}
I am indebted to M. Shaposhnikov for many useful
discussions, and to P. Elmfors, P. Ferreira, K. Kainulainen,
C. Korthals-Altes, G. Moore, T. Prokopec and N. Turok for
conversations or comments. I am grateful to M. Kamionkowski
and the referee for bringing my attention to
\cite{kamturner} and \cite{spokoiny} respectively.

\end{document}